# From Multiparton to Single parton scatterings
# The path towards the Revival of the Transverse Spin

M. MEKHFI[1]

I give a brief historical account on the way I have been conducted to revive the concept of the transverse spin from previous studies on multiparton scatterings.

11.80.La, 12.38.-t, 13.88.+e, 24.70.+s

1- Introduction

The transverse spin has attracted the attention of eminent physicists who pushed it so, that today, transverse spin or more precisely transversity is considered equally important an aspect of nucleon structure as the other two leading-twist spin parton densities: unpolarized and longitudinally polarized . In this paper I will describe, how this new sub-branch of spin physics: The transverse spin , has been launched from early studies of multiparton scattering in QCD. The transverse spin was launched in 1988 in the proceeding of the XIX International Symposium on Multi-hadron Dynamics, Arles (France) [1]. But such work remained unnoticed until 1990 after the publication of the paper *Z. Phys. C45 (1990)* [2] . Since that time, various authors such as R.L.Jaffe [3] and V.Baronne et al [4], to cite only these, manifested desire to give brief comments on what motivated the revival of the transverse spin. But such motivations were nowhere mentioned in the 1990 paper, and even the indication to refer the reader to the 1988 proceeding cited above although it existed in the 1990 paper , was somehow nonchalant and gave the reader the awkward feeling that we got the concept of transverse spin from nothing. To cure this non negligible insufficiency and to make the Z.Phys paper complete (fifteen years later but : *Il n´est jamais trop tard pour bien faire*) I describe to some details the way I have been conducted to re- invent the transverse spin.

---

[1] International Centre for theoretical physics, Trieste, Italy  E-mail: Mekhfi@ictp.it

During seventies and eighties there was a prejudice that transverse spin is small or irrelevant for ultrarelativistic ($m \ll E$) particles, or at least in hard reactions ($m^2 \ll Q^2$ where $Q$ is the deep inelastic momentum scale). This prejudice persisted in spite of large transverse polarizations in $e^+$ and $e^-$ storage rings produced by the Sokolov-Ternov effect. During that period, the physics of transverse spin progressed through isolated works. Studying the Drell-Yan reaction with polarized protons, John P.Ralston and Davison E. Soper introduced the transversity distribution[(1)] $\delta q(x)$ and predicted a double spin asymmetry for transversely polarized baryons [5], but this work was apparently forgotten for about a decade. The reason I believe is that the transverse spin was not their motivation but rather the Drell-Yan mechanism , "An experimental test of these relations would provide a stringent test of the Drell-Yan model" to quote the major and the last sentence of their abstract. Removing the persistent prejudices attached to the transverse spin, stressing the importance transversity would play in the hadron spin structure and putting it at equal footing with conventional polarizations were our contributions to this area of physics and these contributions were behind the resurgence of the transverse spin and its subsequent astonishing development . Now I come to describe the process of thought which led to the transverse spin .As early as 1982 when I was studying the colour and spin structure of multiparton scatterings in quantum chromodynamics (QCD) [6 7 8] ,in the context of my PHD thesis at the international school for advanced studies (ISAS, Trieste, Italy ), I was conducted to investigate the very complex structure of the parton colour in the t-channel. The t-channel being

---

[(1)] I was not aware of the existence of this work at the time I encountered the transverse spin. I was informed about it belatedly

identified as the appropriate channel, had simplified the analysis of the colour structure a great deal. Then applying the same formalism to the spin structure of partons inside the parent hadron, I noticed that, while the parent hadron was unpolarized (or spinless), the spin structure of the multiparton state organized itself in terms of three different contributions: unpolarized, longitudinally polarized and *transversely polarized*. Since the parent hadron was unpolarized, there was no way to favour one spin contribution from the other, hence it appeared that the transverse spin had equal chance to contribute to the hadron spin structure as the other traditional polarizations. In the following I will survey with some details the part of my work on multiparton scatterings which had a close connection to the transverse spin.

2- The multiparton cut amplitude

Composite hadron structure allows the simultaneous participation to hard processes of more than one parton in the hadron. The class of hard disconnected processes at the Born level which are likely to be generated by a disconnected four parton scattering have become of increasing importance since the first observation at the UA1 experiment of 4-jets [9] .Another experiment [10], later followed by the experiment on possible multiparton interactions in photo production at HERA [11] came to confirm the presence of multiparton scatterings. The hadronic part of the squared amplitude Figure 1 is described by a set of double structure functions $\Gamma_{\xi_1\xi_2\xi_2'\xi_1'}(x_1,x_2,\vec{b}_T)$ which are related to the cut amplitude Figure 2. $x_1, x_2$ are the scaled parton momenta and $\vec{b}_T$ is the impact parameter separating the two initial partons. The class of disconnected processes at the Born level enjoys a very important property: connecting gluons (C-gluon), that is, gluons which connect disconnected hard scatterings at the Born level, have negligible contributions at the leading log approximation [67] . As a consequence each hard process evolves separately with the momentum transfer $Q_i^2$ $i=1,2$. This property is the basic ingredient that allows colour and spin correlations to simplify drastically and to suggest the use of t-channel as the appropriate channel to conduct the analysis. More importantly it will allow us to transport our findings in multiparton scatterings to single parton scatterings.

3- Properties of connecting gluons

In dealing with radiative corrections to disconnected multiparton processes, one has to consider two distinct classes of gluons. We call non-connecting gluons those which correct each hard process separately as in single parton processes, and connecting gluons those which connect hard processes $S_1$ and $S_2$ as in Figure 3. This distinction is necessary, for the connecting gluons have the following important and special kinematical constraint.

$$| \sum_{i=C-gluons} \vec{k}_{T_i} |^2 \leq \langle P_T^2 \rangle$$

irrespective of the number of exchanged gluons. $\vec{k}_{T_i}$ is the transverse momentum of the i$^{th}$ C-gluon and $\langle P_T^2 \rangle^{\frac{1}{2}} \sim R^{-1}$ is the average intrinsic transverse momentum inside the hadron of size $R$. This can be seen by inspection of Figure 3 and after making few algebras. This is a consequence of the energy-momentum conservation and of the fact that all external lines in Figure 2 have $k_{T_i} \leq R^{-1}$. This reduction of the available phase space of C-gluons induces a very strong suppression of their contributions at large $Q^2$. I have shown in References 6 and 7 that the leading contribution of C-gluons, denoted $I_C(Q^2)$, comes from collinear gluons and has no logarithmic dependence. Consequently this contribution can be neglected at the leading logarithmic approximation. To order $\alpha_s^n$ we get indeed.

$$\lim_{Q^2 \to \infty} I_{C,n}(Q^2) = cte$$

This theorem guaranties the disconnectedness of the multiparton process to any order in $\alpha_s$ and hence will allow the passage from multiparton to single parton scatterings.

4-The colour structure and the Sudakov suppression of the non-singlets

The key idea to analyse the colour structure which is rather complex is to classify double structure functions according to the colour irreducible representations $\{L\}$ of the diparton system either in the s-channel $(\xi'_1,\xi'_2 \to \xi_1,\xi_2)$ or in the t-channel $(\bar{\xi}_2,\xi'_2 \to \xi_1,\bar{\xi}'_1)$ using the Clebsch-Gordon decomposition. For the example of a diquark system in the proton $(qq)$, the s-channel decomposition is $\{3\} \otimes \{3\} = \{\bar{3}\} \oplus \{6\}$; there corresponds two double structure functions $\Gamma^s_{\bar{3}}$ and $\Gamma^s_6$ where s refer to the s-channel. Colour correlations means that we have $\Gamma^s_{\bar{3}} \neq \Gamma^s_6$. We do have colour correlations at low $Q^2$ where $\Gamma^s_{\bar{3}}$ dominates, since in this regime the baryon is mainly a three-quark colourless state. This correlation is however suppressed à la Sudakov [12] (exponentially). Asymptotic colour non-correlation is most conveniently expressed in terms of t-channel structure functions (L is the t-channel colour representation) and reads

$$\Gamma^t_L(x_1,x_2,\vec{b}_T,Q_1,Q_2) \to 0$$
$$(L \neq 1, Q_1 \text{ or } Q_2 \to \infty)$$

That is, only the singlet component $\Gamma^t_1$ survives the high $Q^2$ regime. Applied to the diparton $(qq)$ system, the t-channel corresponding system is $(q\bar{q})$ with L=1 or 8. $\Gamma^t_1$ survives the evolution while $\Gamma^t_8$ is exponentially suppressed which means that colliding partons lose the colour memory in the cascade.

5- The spin correlation

Now we come to the spin structure which is our major concern. No spin correlations would mean that the density operator in the s-channel has the form[3].

---

[3] This will correspond to singlet states in the t-channel. Nonsinglet states in the t-channel are due to spin correlations and are associated with the transverse spin ( that was true for color, it is in fact a general statement).

$$\Gamma_{\lambda_1 \lambda_2 \lambda_2' \lambda_1'}(x_1, x_2, \vec{b}_T) = \Gamma_1(x_1, x_2, \vec{b}_T)\delta_{\lambda_1 \lambda_1'}\delta_{\lambda_2 \lambda_2'} + \Gamma_2(x_1, x_2, \vec{b}_T)\lambda_1 \lambda_2 \delta_{\lambda_1 \lambda_1'}\delta_{\lambda_2 \lambda_2'} \qquad (1)$$

In other words, if there are no spin correlations, we would need just two structure functions $\Gamma_{1,2}$ to describe hadron polarizations and these are standard unpolarized and longitudinally polarized structure functions. Spin correlations at low $Q^2$ do exist however. In a pion for instance we do have $\lambda_2 = -\lambda_1$. This implies that other terms responsible for correlations are present in (1). Simple manipulation of spin indices shows that such additional terms are associated with the transverse spin. The transverse spin is then present at low $Q^2$. The question one may ask is, do spin correlations vanish asymptotically at high $Q^2$? And if the answer is affirmative, is their fall off exponential as for colour? An exponential fall off of spin correlations would be a signal of the absence of transverse spin at high momentum transfer and hence non observability of it at hadron colliders. Physically we do not expect such spin correlation suppression to occur, because important contributions to the parton cascade are nearly helicity conserving. A careful study of the evolution of the density operator $\Gamma$ confirms the above suggestion. "Diagonalization" of various spin components and their evolutions with the moment transfer, are best performed in the t-channel in analogy with what we did for the colour. Considering $\Gamma$ as an operator in the t-channel we write.

$$\Gamma_{\lambda_1 \lambda_2 \lambda_2' \lambda_1'} = \langle \lambda_1, -\lambda_1' | \Gamma | -\lambda_2, \lambda_2' \rangle$$

One advantage of the t-channel is that helicities of diparton systems $\Delta\lambda_i = \lambda_i - \lambda_i'$ $i=1,2$ are conserved during the evolution, and *this is made possible by the disconnected-ness property at any order of the strong coupling constant (absence of C-gluons[5]) that we have shown previously, otherwise helicity would escape through connecting gluons.* Therefore we choose t-channel structure functions of definite $\Delta\lambda_1$ and $\Delta\lambda_2$. It happens that some of them $|½, -½\rangle$ and $|-½, ½\rangle$ mix during the evolution. In the helicity space, we diagonalize the evolution matrix by a convenient choice of basis, in which the new states $|\Lambda\rangle$ have definite parities.

$$|+1\rangle = |½, ½\rangle \quad |-1\rangle = |-½, -½\rangle$$
$$|0\pm\rangle = \frac{|½, -½\rangle \pm |-½, ½\rangle}{\sqrt{2}}$$

(For gluon –gluon states, all helicities have to be multiplied by 2). States $|0+\rangle$ and $|0-\rangle$ have opposite parities, therefore corresponding structure functions $\Gamma_{\Lambda_1 \Lambda_2}$ no longer mix during the evolution. *The evolution of the double structure functions $\Gamma_{\Lambda_1 \Lambda_2}$ is closely related to the evolution of single parton structure function $\Gamma_{\Lambda_1}^{\Lambda_2}$.* To visualize this, think of legs 2 and 2' of Figure 2 as being a hadron of helicity $\lambda_2$ (delete hadronic legs in mind) which in the t-channel correspond to the $\Lambda_2$ state Figure 4. It is plausible at this point that the states $|\Lambda\rangle = |+1\rangle$, $|\Lambda\rangle = |-1\rangle$ enter the hadronic process with equal footing as the unpolarized and the longitudinally polarized states $|\Lambda\rangle = |0\pm\rangle$, whatever the parton scattering being multiple or single. *The transverse spin is encoded in the t-channel structure functions $\Gamma_{\pm 1 \Lambda_2}$ and $\Gamma_{\pm 2 \Lambda_2}$ respectively for the quark and the gluon and is a result of parton spin correlations inside the parent hadron.* Evolution of transverse spin structure functions showed that spin correlations are slowly suppressed with $Q^2$, thus allowing a possible probe into the spin structure of hadrons at very high $Q^2$, in contrast to colour correlations which undergo an exponential suppression. The following tableau clarifies the correspondence between double and single parton scatterings.

| *Double structure function* | *Single structure function* |
| --- | --- |
| $\Gamma_{0+,\Lambda_2}(x_1, x_2)$ | Unpolarized (quark, gluon) $q(x_1)$; $g(x_1)$ |
| $\Gamma_{0-,\Lambda_2}(x_1, x_2)$ | Longitudinally polarized $q_{+z} - q_{-z}$; $g_{+z} - g_{-z}$ |
| $\Gamma_{+1,\Lambda_2}(x_1,x_2) + \Gamma_{-1,\Lambda_2}(x_1,x_2)$ | x-polarized (quark) $q_{+x} - q_{-x}$ |
| $\Gamma_{+2,\Lambda_2}(x_1,x_2) + \Gamma_{-2,\Lambda_2}(x_1,x_2)$ | Linearly polarized gluon $g_{xx} - g_{yy}$ |

6- Conclusion

In studying the spin and colour structure of multiparton scatterings in QCD, first, I noticed that colour and spin correlations, present at low momentum transfer $Q^2$ implied new multiparton structure functions. These new structure functions correspond, for colour, to t-channel colour non-singlet densities and for spin, to t-channel transverse spin densities . Second, an important theorem in multiparton processes that we had proved, showed that disconnected multiparton scatterings at the Born level remain disconnected at leading logarithmic approximation (absence of connecting gluons). This property was behind the introduction of the t-channel as the appropriate channel to analyze the evolution of the spin structure functions and to make the transition from multiparton to single parton scatterings possible. The T-channel analysis, the evolution of structure functions, in particular the evolution of the transverse spin of interest here and the passage to the transverse spin in single parton scatterings, all these ideas and findings have been presented already at XIX International Symposium on Multiparton Dynamics ( Arles,France,1988) .It is precisely at this stage of the analysis and after the presentation of our results at the Arles Symposium that together with X.Artru I decided to write the Z. Phys.C **45**, 66 (1990) paper dedicated exclusively to single parton scattering which is a summary of what we have learned of the transverse spin from multiparton scatterings to which we added the important section on the observability of the transverse asymmetry in various single parton processes. Later on I added two papers [13] [14] on how transversely polarized experiments may be efficient tools to detect supersymmetry. Let us say, to conclude, that about several years of labors to understand the multiparton spin structure inside hadrons in QCD , were behind the re-invention of the transverse spin in 1990.

**Figure 1**-Example of a disconnected process: x1 and x2 are scaled parton momenta, A and B are unpolarized hadrons.

**Figure 2**-The part of the cut amplitude related to hadron A. $\xi_1, \xi_2, \xi_1', \xi_2',$ are parton color or spin indices .

**Figure 3**-Dashed black lines are non-connecting gluons. Dashed blue lines are connecting gluons. S1 and S2 are hard scatterings.

**Figure 4**-The above spin structure is analogous to that of a single parton scattering in which the parent hadron is in the $\Lambda_2$ state and partons in the $\Lambda_1$ state.

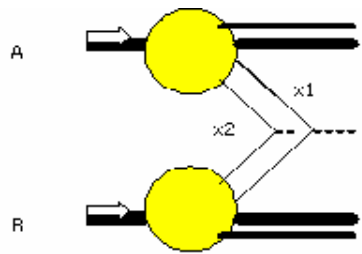

Figure 1

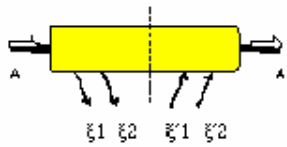

Figure 2

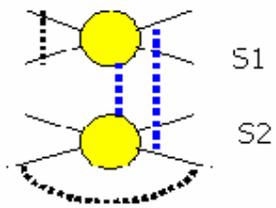

Figure 3

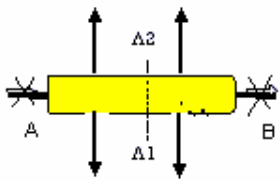

Figure 4

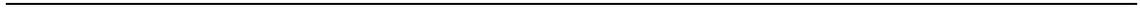